\documentclass[prd,preprint]{revtex4}

	\usepackage{amsmath}%,amssymb} 
	\usepackage{amsfonts}
    \usepackage{amssymb}
	\usepackage{makeidx}
	\usepackage[usenames,dvipsnames]{pstricks}
	\usepackage{epsfig}
	\usepackage{pst-grad} % For gradients
	\usepackage{pst-plot} % F999or axes
    \usepackage{graphicx}
    \usepackage{caption}
	\usepackage{subcaption}
    \usepackage{relsize}

    \usepackage{color}   
    
\newcommand{\be}{\begin{equation}}
\newcommand{\ee}{\end{equation}}
\newcommand{\bea}{\begin{eqnarray}}
\newcommand{\eea}{\end{eqnarray}}

\newcommand{\bsc}[1]{\makebox[0cm]{\vphantom{$\displaystyle #1$}}}

\makeindex

\begin{document}

\title{Reconstructing the metric of the local Universe from number counts observations}
\author{Sergio Andres Vallejo${}^{2,3}$, Antonio Enea Romano${}^{1,2,3}$}

\affiliation{
${}^{1}${Theoretical Physics Department, CERN, CH-1211 Geneva 23, Switzerland} \\
${}^{2}$Instituto de Fisica, Universidad de Antioquia, A.A.1226, Medellin, Colombia \\
${}^{3}${ICRANet, Piazza della Repubblica 10, I--65122 Pescara} \\
%${}^{2}$Yukawa Institute for Theoretical Physics, Kyoto University, Kyoto 606-8502, Japan\\
%${}^{3}$Department of Physics, McGill University, Montr\'eal, QC H3A 2T8, Canada \\
%${}^{3}$Department of Physics and CCTP, University of Crete, Heraklion 711 10, Greece \\
} 

\begin{abstract}
Number counts observations available with new surveys such as the Euclid mission will  be an important source of information about the metric of the Universe. 
We compute the low red-shift expansion for the energy density and the density contrast 
using an exact spherically symmetric solution  in presence of a cosmological constant.
At low red-shift the expansion is more precise than linear perturbation theory prediction. 
We then use the local expansion to reconstruct the metric from the monopole of the density contrast. We test the inversion method using  numerical calculations and find a good agreement within the regime of validity of the red-shift expansion. The method could be applied to observational data to reconstruct the metric of the local Universe with a level of precision  higher than the one achievable using perturbation theory.
\end{abstract}
%\keywords{Cosmological constant, Inflation, Local structure.}

%\pacs{98.80.Es, 98.65.Dx, 98.80.-k}

\maketitle

\section{Introduction}

% At large scales the Universe appears to be isotropic from observations of the cosmic microwave background (CMB) radiation, galaxy number counts and other high redshift observables.

% In particular, the high degree of isotropy observed in the CMB temperature map \cite{Planck:2016ig} points toward this conclusion about our Universe at large scale. This in turns is taken as an experimental footing on which the Universe is assumed to be homogeneous and isotropic at large scales, and therefore can be described by the FLRW solution of Einstein field equations, which hold these symmetries. 
The standard cosmological model is based on the assumption that the Universe is homogeneous and isotropic on sufficiently larges scales, and is confirmed by different observations such as for example the cosmic microwave background (CMB) radiation \cite{Planck:2016ig} or of galaxy catalogues.
However the presence of structure at smaller scales can affect local observations as it was shown in \cite{Romano:2014iea}, and it is therefore important to understand its consequences. The effects of inhomogeneities on cosmological observables have been studied in different cases \cite{Romano:2014iea,Romano:2013kua,Clarkson:2010ej,Romano:2006yc,Romano:2009xw,Ben-Dayan:2014swa,Redlich:2014gga,Romano:2009qx,EneaRomano:2011aa,Marra:2011ct,Romano:2011mx,Romano:2010nc,Krasinski:2014zza,Romano:2012yq,Romano:2012ks,Balcerzak:2013kha,Romano:2012gk,2012GReGr..44..353R,Fanizza:2013doa,RomanoChen:2014,Krasinski:2014lna,Chung:2006xh,Romano:2014tya,Marra:2013rba} such as  dark energy, the luminosity distance \cite{Romano:2010nc,Romano:2012gk,Romano:2011mx} or the expansion scalar \cite{Vallejo:2016} . 
% It has been shown for example that the value of the cosmological constant could be affected significantly by the presence of a local inhomogeneity seeded by primordial curvature perturbations \cite{Romano:2013kua}, which could also lead to the wrong  conclusion of a varying equation of state for dark energy while only a cosmological constant is present \cite{Romano:2010nc}.
These effects are due to the fact that spatial inhomogeneities change the energy of  photons, modifying the cosmological red-shift due to the Universe expansion. As a consequence  some errors are produced in the estimation of parameters based on homogeneous cosmological models.

One important source of information about the Universe are galaxy catalogues since they allow to map the local density field. Since we can only measure the red-shift of  astrophysical objects for which other distance measurement methods such as stellar parallax cannot be applied, it is important to take into account the effects of these inhomogeneities on the metric in order to compute self-consistently the density in red-shift space.
This is particularly important when trying to determine the metric of the Universe. Different numerical \cite{Romano:2013bxa,Celerier:2009sv,Yoo:2008su,Bolejko:2011ys,Chung:2006xh,McClure:2007hy,Mustapha:1998jb} and analytical \cite{Romano:2009mr} inversion methods have been developed in absence of the cosmological constant. More recently a numerical inversion for the luminosity distance  in presence of the cosmological constant was derived in \cite{Tokutake:2016hod}, but no method has been developed to reconstruct the metric from density observation in presence of a cosmological constant. 

In this paper we develop for the first time a low red-shift analytical inversion method in presence of a cosmological constant  to reconstruct the metric from the monopole of the  density contrast, modeling the monopole of local structure with a spherically symmetric exact solution of Einstein fields equations.

The paper is organized as follows. In section II we show how we model the monopole of the local structure using an exact solution of Einstein field equations. In section III we compute  the redshift expansion of the geodesics equations \cite{Vallejo:2016}. In section IV we calculate the density  $\rho(z)$ and the density contrast $\delta(z)$. Finally in section V we develop an analytical method to determine the metric of the Universe from the density contrast.

\section{Modeling the local Universe}
In order to model the monopole component of the local structure we use the LTB solution \cite{Lemaitre:1933,Lemaitre:1933qeLemai,Lemaitre:1931zz,Tolman:1934za,Bondi:1947av} 
\begin{eqnarray}
\label{LTBmetric} %
ds^2 = -dt^2  + \frac{R'(t,r)^2}{1 + 2\,E(r)}dr^2 +R(t,r)^2
d\Omega^2 \,, 
\end{eqnarray}
where $E(r)$ is an arbitrary function of $r$, $R$ is a function of the time coordinate $t$ and the radial
coordinate $r$, and the partial derivative of this function with respect to $r$ is denoted as $R'(t,r)=\partial_rR(t,r)$. 
It follows from the Einstein equations that
\begin{eqnarray}
\label{eq:2} \left({\frac{\dot{R}}{R}}\right)^2&=&\frac{2
E(r)}{R^2}+\frac{2M(r)}{R^3}+\frac{\Lambda}{3} \,,  \\
\label{eq:3} \rho(t,r)&=&\frac{2M'}{R^2 R'} \,, 
\end{eqnarray}
where $M(r)$ is an arbitrary function of the radial cordinate $r$, we adopt a system of units in which $c=8\pi G=1$, and we denote the partial derivative of $R$ with respect to $t$ as $\dot
R=\partial_tR(t,r)$. 

% To compute any quantity in redshift space, in particular $\rho(z)$ and $\delta(z)$, we need to solve the radial null geodesics \cite{Celerier:1999hp} 
% \begin{eqnarray}
% {dr\over dz}&=&{\sqrt{1+2E(r(z))}\over {(1+z)\dot {R'}[t(z),r(z)]}} \,,
% \label{eq:4} \\
% {dt\over dz}&=&-\,{R'[t(z),r(z)]\over {(1+z)\dot {R'}[t(z),r(z)]}} \,. 
% \label{eq:5} 
% \end{eqnarray}

The analytical solution of eq.(\ref{eq:2}) can be derived \cite{Zecca:2013wda,Edwards01081972} if we introduce new functions $\rho_0(r)$ and $k(r)$, and a new coordinate $\eta=\eta(t,r)$ given by
\bea
\frac{\partial \eta}{\partial t}|_r&=& \frac{r}{R}=\frac{1}{a}\,, \label{eq:6}\\
\rho_0(r)&=&\frac{6 M(r)}{r^3}\,, \label{eq:7}\\
k(r)&=&-\frac{2E(r)}{r^2}\,. \label{eq:8}
\eea
Without any loss of generality we will adopt the coordinate system in which $\rho_0(r)$ is a constant, which is known as the FLRW gauge. 
Then we can express eq.(\ref{eq:2}) in the form
\begin{equation}
\left(\frac{\partial a}{\partial \eta}\right)^2= -k(r)a^2 + \frac{\rho_0}{3}a + \frac{\Lambda}{3}a^4 \,. \label{eq:9}
\end{equation}
 
% The coordinate $\eta$ is defined implicitly by eq.(\ref{eq:6}), and can be considered a generalization of the conformal time in a FLRW Universe. Therefore if we integrate this equation we can find the relation between $\eta$ and $t$ given by \cite{RomanoChen:2014}
% \begin{equation}
% t(\eta,r)=\displaystyle\int_{0}^{\eta} a(x,r) \, dx +t_b(r) \,, \label{eq:10}
% \end{equation}
% where $t_b(r)$ is a functional constant of integration known as the bang function, because it corresponds to the fact that the scalar factor in this solution can vanish at different times at different radius. We will just consider models in which the scale factor vanishes at the same time at all radius, ie. $t_b(r)=0$.
The solution to this equation can be written in the form \cite{EneaRomano:2011aa}
\begin{equation}
a(\eta,r)=\frac{\rho_0}{k(r)+3 \wp (\frac{\eta}{2};g_2(r),g_3(r))} \,. \label{aeta} 
\end{equation}
where $\wp(x;g_2,g_3)$ is the Weierstrass elliptic function and
\bea
g_2(r)=\frac{4}{3}k(r)^2 \,,\quad
g_3(r)=\frac{4}{27} \left(2k(r)^3 -\Lambda\rho_0^2\right)\,. \label{eq:12}
\eea
The relation between $t$ and $\eta$ can be found by integrating eq.(\ref{eq:6}) and is given by \cite{Vallejo:2016}
\begin{equation}
t(\eta,r)=\frac{2 \rho _0}{3 \wp '\left(\wp
   ^{-1} \left( {-\frac{k(r)}{3}} \right)\right)} \left[ \ln \left(\frac{\sigma \left(\frac{\eta
   }{2}-\wp ^{-1}\left(-\frac{k(r)}{3}\right)\right)}{\sigma \left(\frac{\eta}{2}+\wp ^{-1}\left(-\frac{k(r)}{3}\right)\right)}\right)+\eta  \zeta \left(\wp
   ^{-1}\left(-\frac{k(r)}{3}\right)\right) \right] \,, \label{eq:21}
\end{equation}
where $\wp '$ is the derivative of the Weierstrass' elliptic function $\wp$.
% $\wp^{-1}$, $\zeta$ and $\sigma$ are defined according to the following relations 
% \begin{align}
% \wp^{-1}\left(\wp\left(x\right)\right) &= x \,, \label{eq:22} \\ 
% \zeta ' \left(x\right) &= - \wp \left(x \right) \,, \label{eq:23} \\
% \frac{\sigma ' \left(x \right)}{\sigma \left(x \right)} &= \zeta \left(x \right) \,. \label{eq:24}
% \end{align}

In terms of the new coordinate $\eta$ and the function $a(\eta,r)$ the radial null geodesic equations are given by \cite{Romano:2009xw}
\bea
\frac{d\eta}{dz}&=&-\frac{\partial_r t(\eta,r) + G(\eta,r)}{(1+z)\partial_\eta G(\eta,r)} \,, \label{eq:13} \\
\frac{dr}{dz}&=&\frac{a(\eta,r)}{(1+z)\partial_\eta G(\eta,r)} \,, \label{eq:14}
\eea
where the function $G(\eta,r)$ has an explicit analytical form given by
\begin{align}
G(\eta,r)&\equiv\frac{R'(t(\eta,r),r)}{\sqrt{1-k(r)r^2}} \, , \label{eq:15} \\
 R'(t(\eta,r),r)&=\partial_r (a(\eta,r)r) - a^{-1} \partial_\eta (a(\eta,r)r) \partial_r t(\eta,r)\,. \label{eq:16}
\end{align}

Finally, in terms of the new coordinate system the density profile is given by
\bea
\rho(\eta,r)&=& \rho(t(\eta,r),r) = \frac{\rho_0}{a(\eta,r)^2 R'(t(\eta,r),r)} \, . \label{eq:17}
\eea
% where $R'(t(\eta,r),r)$ is given in eq.(\ref{eq:16}).
\section{Analytical approximations}
In order to obtain a low-redshift formula for the density contrast and the density profile we expand the function $k(r)$ as
\begin{align}
k(r)&=k_0 + k_1 r + k_2 r^{2} + ... \,, \label{eq:18}
\end{align}
From the exact solution for $a(\eta,r)$ we can obtain an expansion for $t(\eta,r)$ according to
\begin{equation}
t(\eta,r)=t_0(r)+a(\eta_0,r)(\eta-\eta_0)+\frac{1}{2}\partial_{\eta}a(\eta_0,r)(\eta-\eta_o)^2+... \,,\label{eq:19}
\end{equation}
where we defined $t_0(r)$ by
\begin{equation}
t_0(r)\equiv t(\eta_0,r) \,. \label{eq:20}
\end{equation}

%\textcolor{red}{Now}.
The low red-shift Taylor expansion for the geodesic equations can be found in \cite{Romano:2010nc}.
%The use of the exact expression for $t(\eta,r)$ improves the accuracy for the expansion for the geodesics respect to previous calculations \cite{Romano:2012gk}, which were based  on a perturbative expansion of $t_0(r)$, rather than the use of the exact value.\\
However in order to reconstruct correctly the metric all the quantities which depend on the coefficients of $k(r)$ appearing in the analytical approximations, such as $t_0'(0)$ and $t_0''(0)$, must be written  explicitly in terms of those coefficients. Therefore we introduce the dimensionless quantities $K_n$, given by $k_n= K_n (a_0 H_0)^{n+2}$, and applying the chain rule for the derivatives of $t_0(r)$ it follows from eq.(\ref{eq:21}) and (\ref{eq:20}) that
\bea
t_0'(0)&=&\frac{\partial t_0(r) }{\partial r}|_{r=0}=\left(\frac{\partial t_0 }{\partial k}\frac{\partial k}{\partial r}\right)|_{r=0}= a_0 \alpha K_1 \,, \label{eq:25} \\
t_0''(0)&=&  a_0 (a_0 H_0) \left( \beta K_{1}^{2} + 2 \alpha K_2\right)  \, , \label{eq:26}
\eea 
where $\alpha$ and $\beta$ are dimensionless quantities given by
\bea
\alpha&=&\frac{(a_0 H_0)^{3}}{a_0}\frac{\partial t_0 }{\partial k}|_{k=k_0} \,, \label{eq:27}\\
\beta&=& \frac{(a_0 H_0)^{5}}{a_0}\frac{\partial^{2} t_0 }{\partial k^{2}}|_{k=k_0} \, . \label{eq:28}
\eea 
Now we can replace the expressions from eq.(\ref{eq:25}) and (\ref{eq:26}) into the low-redshift Taylor expansion for the geodesic equations given in \cite{Vallejo:2016}. 
% In order to do this the solution of the geodesic equations is expanded according to
% \bea
% r(z)&=&r_1z+r_2z^2+r_3z^3 + ... \label{eq:29}\\ 
% \eta(z)&=&\eta_0 +\eta_1 z +\eta_2 z^2+... \label{eq:30}
% \eea
% The solution of the system of differential equations given by the geodesic equations can be mapped into the solution of a system of algebraic equations for the coefficients of the expansions shown in eq.(\ref{eq:29}) and (\ref{eq:30}) after substituting these expansions in eq.(\ref{eq:13}) and (\ref{eq:14}) . 
%The general expressions for the coefficients are rather  complicated, and are  given in details in appendix B.
We will consider the case in which $k_{0}=0$, which is enough to understand qualitatively the effects of the inhomogeneity, since this term corresponds to the homogeneous component of the curvature function $k(r)$. The results for the general case for not vanishing $k_0$ are given in Appendix B.
% Indeed when there is no inhomogeneity $k_0$ is just the curvature of a $FLRW$ solution and therefore is not related to any new physical effect not already known from the standard cosmological model. 

The expansion for $r(z)$ and $\eta(z)$ requires to first expand the geodesics equations and then to solve a complicated system of linear equations. We have used the Mathematica software for all the analytical calculations, writing some  simplifying routines to express all the results in terms of $H_0$ and other dimensionless parameters given below. This ensures an immediate check of the dimensional correctness of the results and facilitate  their physical interpretation.  The simplifying routines allow to eliminate the elliptic functions and their derivatives according to the procedure explained in Appendix A. Without this automatic manipulations the formulae would be very cumbersome and difficult to derive.
For the geodesics redshift expansion we get the  coefficients 
\begin{align}
\eta _1 &= -\frac{\alpha  K_1+1}{a_0 H_0} \,, \label{eq:31}\\
\eta _2 &= \frac{1}{12 a_0 H_0 \Omega _M} \bsc{\int A} \left[ 2 K_1^2 \left(-4 \alpha +9 \alpha ^2 \left(\Omega _M-1\right) \Omega _M-3 \beta  \Omega _M\right)+K_1 \left(27 \alpha  \Omega _M^2-12 \alpha  \Omega _M-4\right) \right. + \nonumber \\ & \quad {}  \left. \bsc{\int A}   +3 \Omega _M \left(3 \Omega _M-4
   \alpha  K_2\right) \right]  \,, \label{eq:32} \\
r_1 &= \frac{1}{a_0 H_0} \,,  \label{eq:33}\\
r_2 &= \frac{K_1 \left(-18 \alpha  \Omega _M^2+12 \alpha  \Omega _M+4\right)-9 \Omega _M^2}{12 a_0 H_0 \Omega _M} \, , \label{eq:34}\\ 
r_3 &= \frac{1}{72 a_0 H_0 \Omega _{\Lambda } \Omega _M^2} \bsc{\int A} \left[2 K_1^2 \left(4 \zeta _0+\Omega _{\Lambda } \left(-27 \left(8 \alpha ^2+\beta \right) \Omega _M^3+18 \left(3 \alpha ^2-4 \alpha +\beta \right) \Omega _M^2+ \right. \right. \right.  \nonumber \\ & \quad {} \left. \left. \left. \bsc{\left( \left( \left( A \right) \right) \right)} +162 \alpha ^2 \Omega _M^4+36 \alpha 
   \Omega _M+4\right)-6 \zeta _0 \Omega _M+2 \Omega _M\right)+12 K_1 \Omega _{\Lambda } \Omega _M^2 \left(27 \alpha  \Omega _M^2-24 \alpha  \Omega _M-5\right)+ \right.  \nonumber \\ & \quad {} \left. \bsc{ \left( \left( \left( \left( \left(   A  \right) \right) \right) \right) \right) } +3 \Omega _{\Lambda } \Omega _M
   \left(K_2 \left(-36 \alpha  \Omega _M^2+24 \alpha  \Omega _M+8\right)+3 \left(9 \Omega _M-4\right) \Omega _M^2\right) \right] \,. \label{eq:35}
\end{align}
It is important to observe that all these formulae have the correct dimensions, since all the relevant quantities have been expressed in  dimensionless form,  apart from the dimensionfull prefactor $H_0^{-1}$. In the above equations we have introduced the parameters $a_0$, $H_0$, $\Omega_M$, $\Omega_{\Lambda}$, $T_0$ and $\zeta_0$ according to their corresponding definitions given in  \cite{EneaRomano:2011aa,Vallejo:2016}. 
% \bea
% \rho_0&=&3 \Omega_M a_0^3 H_0^2 \,, \label{eq:36} \\
% \Lambda&=&  3 \Omega_{\Lambda} H_0^2 \,, \label{eq:37}\\
% T_0&=& \eta_0 \left(a_0 H_0 \right) \,, \label{eq:38} \\
% a_0 & = &a (\eta_0,0) \,, \label{eq:39}\\
% H_0 & = & H(\eta_0,0) \,, \label{eq:40}\\
% \zeta_0 & = & \zeta \left(\frac{T_0}{2};\frac{4 K_0^2}{3},\frac{4}{27}\left(2 K_0^3 - 27 \Omega_{\Lambda} \Omega_M^2\right)\right) \,, \label{eq:41}
% \eea
% where $\zeta$ is the Weierstrass Zeta Function.
As can be seen from these expressions the effects of the inhomogeneity start to show respectively at first order for $\eta(z)$ and second order for $r(z)$ as can be seen from the formluae we found.

\section{Formulae for the density profile and density contrast at low redshift}
The density profile in redshift space is given by 
\begin{equation}
\rho(z) = \rho (\eta(z),r(z)),  \label{eq:42}
\end{equation}
and substituting the formulae for $\eta(z), r(z)$ we obtain after expanding to second order in red-shift %\textcolor{blue}{using the Wolfram Mathematica software}
\begin{align}
\rho(z) &= 3 \Omega_M H_0^{2} + \rho_1 z + \rho_2 z^2 \, , \label{eq:43}\\
\rho_1 &= H_0^2 \left(4 K_1 \left(3 \alpha  \Omega _M+1\right)+9 \Omega _M\right) \, , \label{eq:44} \\
\rho_2 &= -\frac{H_0^2}{12 \Omega _{\Lambda } \Omega _M} \left[K_1^2 \left(\Omega _{\Lambda } \left(-18 \left(25 \alpha ^2-4 \alpha +5 \beta \right) \Omega _M^2+81 \alpha ^2 \Omega _M^3-300 \alpha  \Omega _M-40\right)+ \right. \right. \nonumber \\ & \quad {} \left. \left. \bsc{\left( \left( \left( \left( A \right) \right) \right) \right)} +20 \left(\Omega _M-\zeta
   _0\right)\right)+18 K_1 \Omega _{\Lambda } \Omega _M \left(3 \alpha  \Omega _M^2+(2-24 \alpha ) \Omega _M-8\right)+ \right. \nonumber \\ & \quad {}  \left. \bsc{\left( \left( \left( \left( \left( A \right) \right) \right) \right) \right)} -12 \Omega _{\Lambda } \Omega _M \left(5 K_2 \left(3 \alpha  \Omega _M+1\right)+9
   \Omega _M\right)\right] \, . \label{eq:45}
\end{align}
It is important to observe that all these formulae have the correct dimensions, since all the relevant quantities have been expressed in  dimensionless form, apart from the dimensionfull prefactor $H_0^2$. Another important result is that $a_0$ does not appear anywhere, as  expected, since the value of $a_0$ is arbitrary and physically observable quantities should not depend on it.

In order to derive an analytical approximation for the density contrast  we must first define what is the background density.
For any scalar $\Phi(t,r)$ we first define the sub-horizon volume average on constant time slices  
\bea
\overline{\Phi}(t) &=& \frac{\int \Phi(t,r) dV(t)}{\int dV(t)}  \label{dV} \,, \\ %} = \mathlarger{\frac{\mathlarger{\int_0^{r_{Hor}}} \rho(t,r) \frac{R(t,r)^2 R'(t,r)}{\sqrt{1-k(r)r^2}} dr}{\mathlarger{\int_0^{r_{Hor}}} \frac{R(t,r)^2 R'(t,r)}{\sqrt{1-k(r)r^2}} dr}} \, . \label{eq:46} \\
\int dV(t) &=&\int_0^{r_{Hor}(t)} \frac{R(t,r)^2 R'(t,r)}{\sqrt{1-k(r)r^2}} dr 
\eea
where the upper limit of the integral $r_{Hor}(t)$ is the comoving horizon as a function of time, and determines the region of space causally connected with the central observer at time $t$. Note in fact that spatial averaging on super-horizon scales is not physically meaningful, since the effects of super-horizon  structures are unobservable \cite{Romano:2006yc}.  

We  can then evaluate  $\overline{\Phi}(t)$ at the time $t(z)$ corresponding to a given redshift $z$, i.e. the time along  null radial geodesics, and define the background value of $\Phi$ at redshift $z$ as 
\bea
\overline{\Phi}(z)&=&\overline{\Phi}(t(z)) \label{} \,.
\eea
Applying this definition of background value to $\rho$ we can define the density contrast
\bea
\delta(z)&=& \frac{\rho(z)}{\overline{\rho}(z)} -1 \,.  \label{dcLTB}
\eea

% In order to model the local structure in our Universe we will assume that the background Universe corresponds to a homogeneous FLRW solution of Einstein equations, this is motivated by the fact that observations at high redshift point towards a highly isotropic, and therefore homogeneous, Universe. T
If the size of the local inhomogeneity is sufficiently smaller than the volume over which the integral in eq.(\ref{dV}) is performed then $\overline{\rho}$ will get most of its contribution from the asymptotically homogeneous region and the average density will be well approximated by the asymptotic density
%For this type of compensated inhomogeneities the asymptotic behavior of the function $k(r)$ allows to define background quantities  as
\bea
%\overline{\rho}(t) = \lim_{r\to\infty}\rho(t,r) \, , \\ %\qquad \text{or equivalently} \qquad 
\overline{\rho}(z) = \lim_{r\to\infty} \rho(t(z),r)  \label{rhoinf} \,.
%\overline{H}(z) = \lim_{r\to\infty} H(t(z),r) \, , \label{eq:59} \,.
\eea

This is clearly true for compensated structures of any size, but it also applies to uncompensated  structures whose size is sufficiently smaller than the asymptotic homogeneous region.

We can now re-write the background energy density as
\bea
\overline{\rho}(z)&=& 3 (H_0^b)^2 \Omega_M^b (1+z)^3 , \label{eq:48} \\
H_0^b&=&\overline{H}(0) \,, \\
\Omega_M^b&=& \frac{\overline{\rho}(0)}{3 (H_0^b)^2} \,, \\
\Omega_{\Lambda}^b&=&1-\Omega_M^b \,,
\eea
where the upper-script $^b$ stands for background and  $H$ is the expansion scalar $H(t,r)$ \cite{Vallejo:2016}.
Note that as a consistency check we have verified numerically that, as expected, for the  central sub-horizon inhomogeneities we considered in this paper  both $\overline{H}(z)$ and $\overline{\rho}(z)$ computed through  the volume average defined in eq.(\ref{dV}) are in good agreement with eq.(\ref{rhoinf}).

% We can then define the background parameters  $H_0^b$ and $\Omega_M^b$ as
% \begin{equation}
% H_0^b = \overline{H}(0) \, , \qquad \text{and}  \qquad \Omega_M^b= \frac{\overline{\rho}(0)}{3 (H_0^b)^2} \, . \label{eq:58}
% \end{equation}

%In this paper we will consider compensated structure when comparing to numerical results, but the formulae we derive can also be applied to uncompensated structures  when the conditions explained above are satisfied.

% As can be seen, this background density corresponds to the matter energy density in a homogeneous Universe in presence of a cosmological constant. 
Expanding the density contrast we find 
\begin{align}
\delta(z) &= \delta_0 + \delta_1 z + \delta_2 z^2 \, , \label{eq:49}\\
\delta_0 &= \left(\frac{H_0}{H_{0}^b}\right)^2 \frac{ \Omega _M}{\Omega _{M}^b} -1  \, , \label{eq:50}\\ 
\delta_1 &= \left(\frac{H_0}{H_{0}^b}\right)^2 \frac{4  K_1 \left(3 \alpha  \Omega _M+1\right)}{3  \Omega _{M}^b} \, , \label{eq:51} \\
\delta_2 &= - \left(\frac{H_0}{H_{0}^b}\right)^2 \frac{1}{36  \Omega _{\Lambda } \Omega _M
   \Omega _{M}^b}  \Bigg[ 18 K_1 \Omega _{\Lambda } \Omega _M^2 \left(3 \alpha  \Omega
   _M+2\right)-60 K_2 \Omega _{\Lambda } \Omega _M \left(3 \alpha  \Omega _M+1\right)+  \nonumber \\ & \quad {}   +K_1^2 \left(\Omega _{\Lambda } \left(-18 \left(25 \alpha ^2-4 \alpha +5 \beta \right) \Omega _M^2+81 \alpha ^2 \Omega
   _M^3-300 \alpha  \Omega _M-40\right)+20 \left(\Omega _M-\zeta _0\right)\right)\Bigg] \, . \label{eq:52}
\end{align}
It is easy to check  that all these formulae have the correct dimensions because all the relevant quantities have been expressed in  dimensionless form.

As can be seen the first order coefficients of the expansion of $\rho(z)$ and $\delta(z)$ depend on $K_1$, while the second order depend on both $K_1$ and $K_2$.

The procedure to reduce the analytical formulae to this form is rather complicated since it involves to express wherever possible all the intermediate expressions in terms of physically meaningful quantities and  we give more details about it in  appendix A. The formulae for the case in which $K_0$ is different from zero are rather cumbersome and we give them in appendix B.
 
\subsection{Comparison with numerical and pertubative calculations}

\begin{figure}[ht]
    \begin{subfigure}{0.48\textwidth}
       \includegraphics[width=\textwidth]{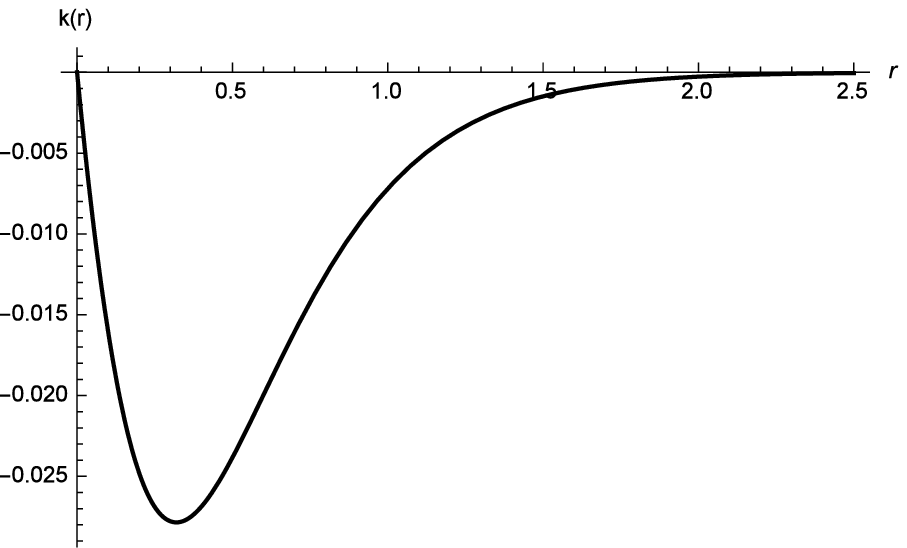}
    \end{subfigure}
    \quad
	\begin{subfigure}{0.48\textwidth}
       \includegraphics[width=\textwidth]{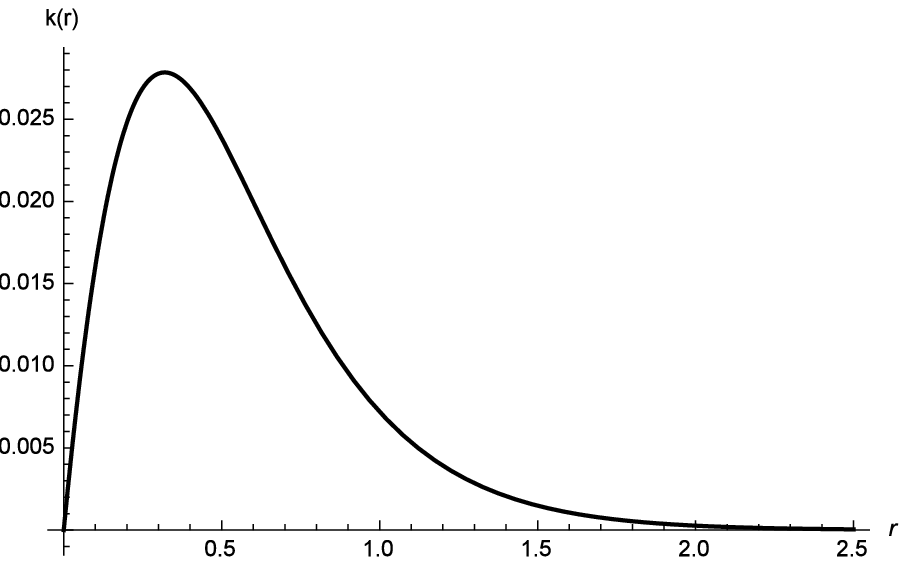}
    \end{subfigure}
    
 \caption{The function $k(r)$ defined in  eq.(\ref{eq:57}) is plotted in units of  $H_0^{2}$  as a function of the radial coordinate in units of $H_0^{-1}$.}   
\label{knp}
\end{figure}

In order to test the formulae we have derived in the previous section we consider inhomogeneities defined by a spatial curvature function $k(r)$ of  this type
\begin{equation}
k(r) = \pm \frac{r}{5}  [1-\tanh (2 r)], \label{eq:57}
\end{equation}
which is plotted in fig.(\ref{knp}). 
We solve numerically eq.(\ref{eq:2}) and the radial null geodesic equations given in \cite{Celerier:1999hp}. This type of function $k(r)$ is satisfying the assumption we made in the previous section that $k(0)=k_0=0$ and it corresponds to compensated structures making it easy to define background quantities according to eq.(\ref{rhoinf}). 

For the models we consider in this section we have $H_0^b=H_0$ and $\Omega_M^b=\Omega_M$.
%This same approach can be used to get $\overline{H}(z)$ and even the background scale factor $\overline{a}(t)$ from the function $a(t,r)=\frac{R(t,r)}{r}$. 
It also follows for these models that the curvature of the background solution is $k^b=\lim_{r\to\infty}k(r)=0$, which corresponds to a flat homogeneous Universe. 
%In order to find the numerical solutions we consider models with $H_0=1$, $a_0=1$, and $\Omega_{\Lambda}=0.692$. 
% Then we solved numerically eq.(\ref{eq:2}) for the radial interval $r\in(0,2.5)$ and the time interval $t\in(0,2)$. And we also solved eq.(\ref{eq:4}) and (\ref{eq:5}) for the redshift interval $z\in(0,0.2)$. 

\begin{figure}[ht]
 \centering
    \begin{subfigure}{0.48\textwidth}
       \includegraphics[width=\textwidth]{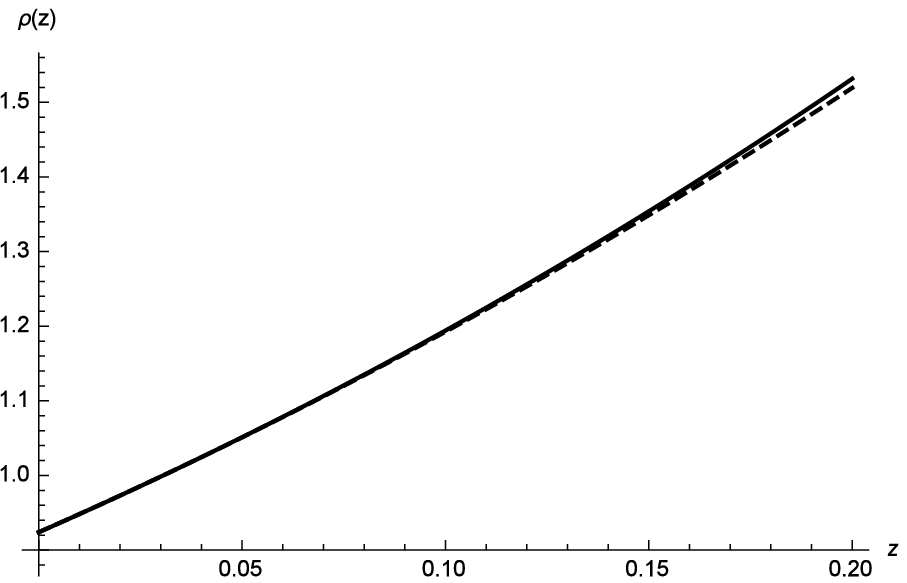}
    \end{subfigure}
    \quad
	\begin{subfigure}{0.48\textwidth}
       \includegraphics[width=\textwidth]{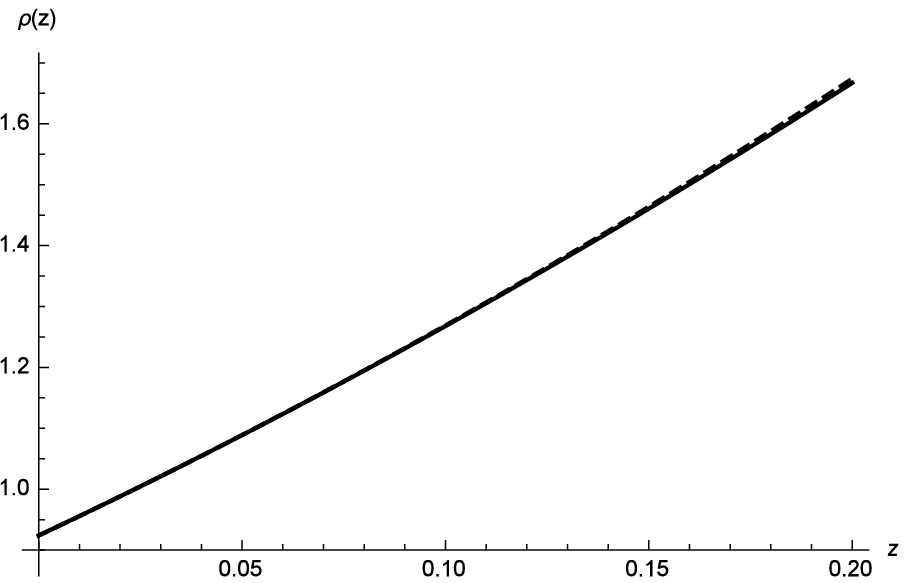}
    \end{subfigure}
 \caption{The density profile in units of $H_0^{2}$ is plotted as a function of redshift. The left and right are plots are for the inhomogeneities corresponding to Fig. 1. The solid lines are for the numerical calculation and the dashed lines for  the analytical approximation.}   
\label{rhonp}
\end{figure}

\begin{figure}[ht]
 \centering
    \begin{subfigure}{0.48\textwidth}
       \includegraphics[width=\textwidth]{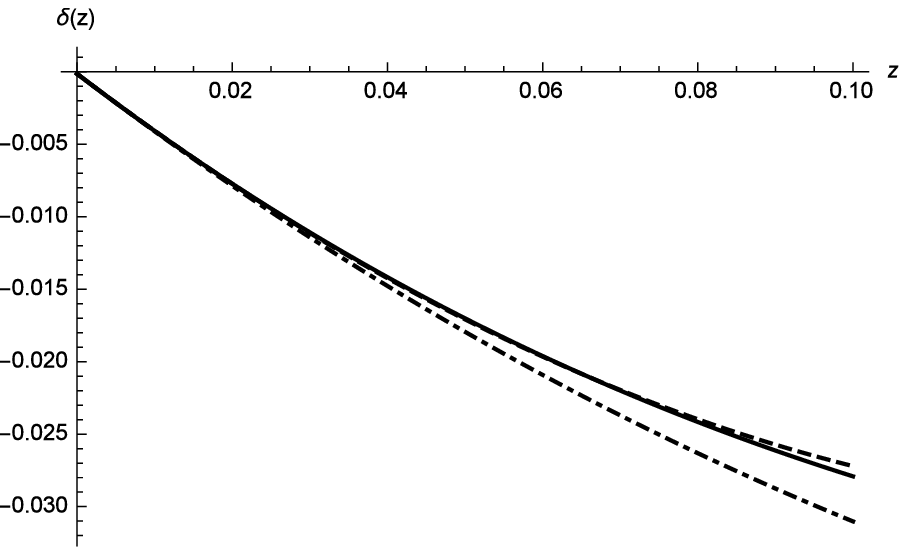}
    \end{subfigure}
    \quad
	\begin{subfigure}{0.48\textwidth}
       \includegraphics[width=\textwidth]{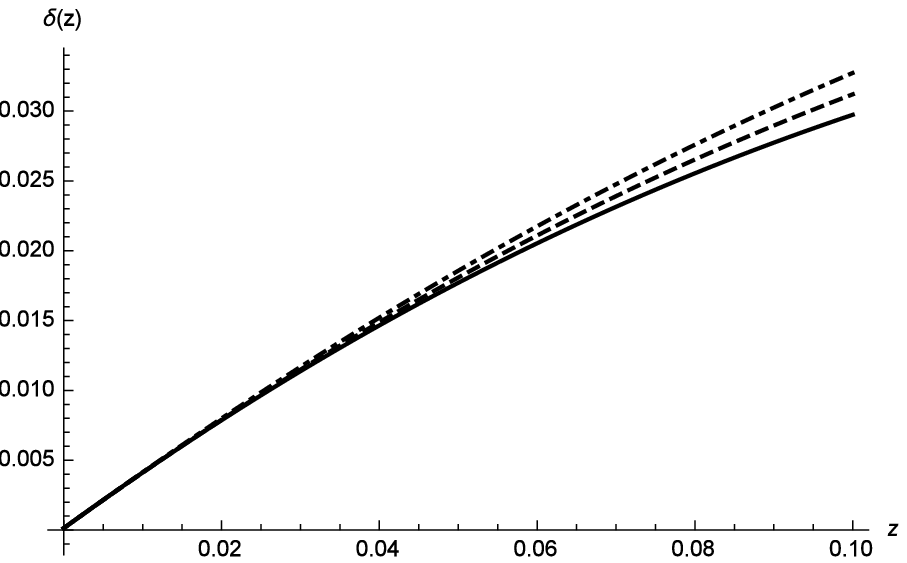}
    \end{subfigure}
 \caption{The density contrast is plotted as a function of redshift. The left and right  plots are for the inhomogeneities corresponding to Fig. 1. The solid lines correspond to the numerical solution, the dashed lines to the analytical formula we derived and the dot-dashed lines to the perturbation theory result.}   
\label{deltarhonp}
\end{figure}

% We then plotted the density profile of the inhomogenity for both the numerical solution and the analytical approximation we derived up to a redshift of $z_l=0.1$. 
In order to compare our results to linear perturbation theory we  compute the perturbation theory prediction for $\delta(z)$ according to \cite{Turner:1992}
\begin{equation}
\delta (z) \approx - 3 \delta H(z) (\Omega_M^b)^{-0.55} \, . \label{eq:60}
\end{equation}
 As can be seen in fig.(\ref{deltarhonp}) and fig.(\ref{rhonp}) at low red-shift the analytical formulae for $\rho(z)$ and $\delta(z)$ derived in eq.(\ref{eq:43}) and (\ref{eq:49}) are in good agreement with the numerical calculations and are more accurate than the perturbation theory prediction in eq.(\ref{eq:60}).

% With the goal of comparing more precisely our analytical approximation with the perturbative formula in eq. (\ref{eq:60}) we plot the relative percentual difference with respect to the numerical solution for both models as can be seen in Fig. 4. 
\begin{figure}[ht]
 \centering
    \begin{subfigure}{0.48\textwidth}
       \includegraphics[width=\textwidth]{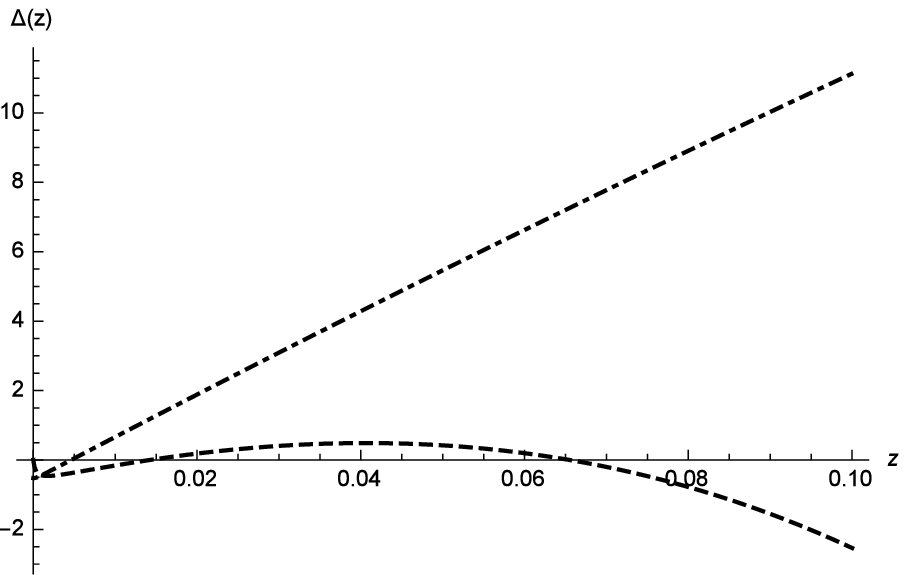}
    \end{subfigure}
    \quad
	\begin{subfigure}{0.48\textwidth}
       \includegraphics[width=\textwidth]{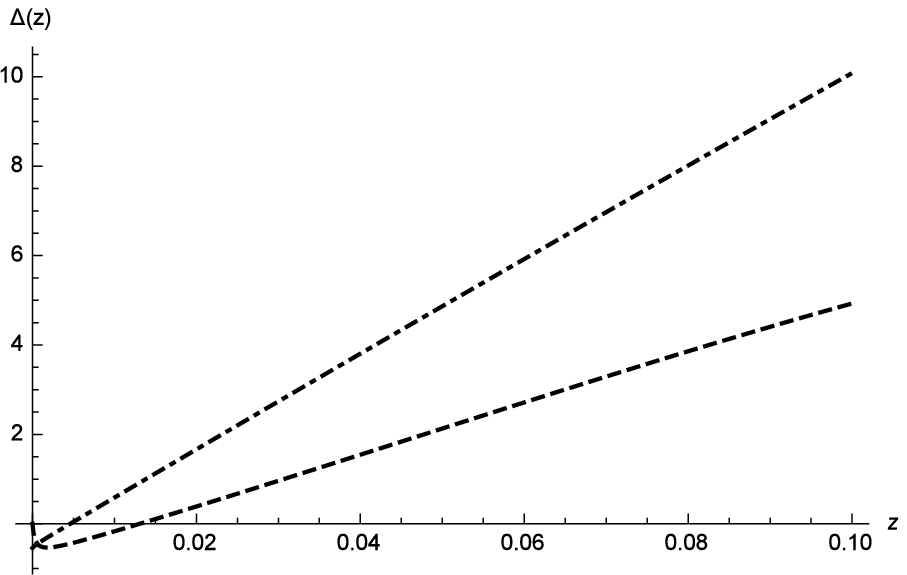}
    \end{subfigure}
 \caption{The relative percentual difference between different analytical approximations and the numerical calculation $\Delta(z)=100 \left(\frac{\delta^A(z)}{\delta^N(z)}-1\right)$  is plotted as a function of redshift. The left and right plots  correspond to inhomogeneities in Fig. 1. The dashed lines are for the analytical formula and the dot-dashed lines for the perturbation theory approximation.}   
\label{Deltanp}
\end{figure}

% It is well known from cosmological perturbation theory  \cite{Malik:2009cp} that the density perturbation in the comoving  gauge corresponds to a gauge invariant quantity, called comoving density contrast. Since in the LTB solution is written in the comoving gauge, it follows that the density contrast defined in eq.(\ref{dcLTB}) is gauge invariant, and corresponds to a physically observable quantity. 

% The comoving density contrast appears also in the cosmological perturbation equations written in the Newton gauge:
% \bea

% \eea
% and it can consequently be used to find the relation between the two gauges if necessary.

\section{Reconstruction of the metric from the density contrast}
% In order to reconstruct the monopole component of the metric %from the monopole component of the density contrast 
% we will assume that at low redshift the density contrast can be expressed according to equation (\ref{eq:48}). 
In the previous section we have obtained the red-shift expansion of the density contrast in terms of the dimensionless coefficients $K_i$, and we will now use this to solve the inversion problem, i.e. to obtain  $K_i$ from the density contrast.
Note that in the coordinates we are using the coefficients $K_i$ complete determine the metric, so that we will be able to reconstruct the metric from the density contrast.

% We can solve the system of equations (\ref{eq:49})-(\ref{eq:51}) for the coefficients $k_1$ and $k_2$ in terms of the density contrast. 
% This is equivalent to finding the Bardeen potential in cosmological perturbation theory in the Newtonian-longitudinal gauge from the density contrast given, and under the right assumptions finding both must be the same thing. 

%From eq.(\ref{eq:50}) we get 
%\begin{equation}
% \Omega_M = \Omega_M^h  \left(1+\delta _0\right) \left(\frac{H_0^h}{H_0}\right)^2 \, , \quad \text{or} \qquad 
%\Omega_M^b = \frac{\Omega_M}{\left(1+\delta _0\right)} \left(\frac{H_0}{(H_0^b)}\right)^2 \, , \label{eq:53}
%\end{equation}

% and after replacing it in 

From eq.(\ref{eq:51}) and (\ref{eq:52}) we can  solve the system of equations for the coefficient of the $k(r)$ expansion $K_1$ and $K_2$   
\begin{align}
k(r) & \approx K_1 (a_0 H_0)^3 r + K_2 (a_0 H_0)^4  r^2 \, , \label{eq:54} \\ 
K_1 &= \left(\frac{H_{0}^b}{H_{0}}\right)^2 \frac{3 \Omega _{M}^b \delta _1 }{4  \left(3 \alpha  \Omega _M+1\right)} \, , \label{eq:55} \\
K_2 &= \left(\frac{H_{0}^b}{H_{0}}\right)^4 \frac{3 \Omega _{M}^b}{320\Omega _{\Lambda } \Omega _M \left(3 \alpha  \Omega
   _M+1\right){}^3} \Bigg[8 \Omega _{\Lambda } \Omega _M \left(3 \alpha  \Omega _M+1\right) \left(9 \alpha  \delta _1 \Omega _M^2+6
   \left(4 \alpha  \delta _2+\delta _1\right) \Omega _M +   \right. \nonumber \\ & \quad {}  \left. \bsc{\left( \left( A \right) \right)} + 8 \delta _2 \right)+\delta _1^2  \Omega _{M}^b  \left(\frac{H_{0}}{H_{0}^b}\right)^2\left(\Omega _{\Lambda } \left(-18 \left(25
   \alpha ^2-4 \alpha +5 \beta \right) \Omega _M^2+81 \alpha ^2 \Omega _M^3-300 \alpha  \Omega _M-40\right)+  \right. \nonumber \\ & \quad {}  \left. \bsc{\left(  \frac{A}{B} \right)} +20 \left(\Omega _M-\zeta
   _0\right)\right)\Bigg] \, . \label{eq:56}
\end{align}

It can be easily checked that all these formulae have the correct dimensions, since all the relevant quantities have been expressed in  dimensionless form.

The linear coefficient $K_1$ depends on  $\delta_1$, while the second order coefficient $K_2$ depends  on both  $\delta_1$ and $\delta_2$. This is naturally expected since a homogeneous Universe corresponds to $K_1=K_2=\delta_1=\delta_2=0$. As a consistency check it can be easily verified that in fact $K_1=K_2=0$ in the  homogeneous limit, i.e. when $\delta_1=\delta_2=0$.
% The divergence for $\delta_0=-1$ is just an apparent singularity related to the system of coordinates in which we are performing the calculations. In a 
% It is interesting to note that the solution diverges when there is a supervoid at the center of the inhomogeneous local Universe, i.e. $\delta_0=-1$, as can be seen from the expressions above. This may be arising because of an apparent singularity given our coordinate choice to solve the problem. 

It is important to note that the nonlinearity of Einsteins equations implies that the solution of the inversion problem (IP) is not unique. The input of the IP is in fact the monopole $\delta(z)$ of the density contrast, but the metric obtained applying to $\delta(z)$ the inversion is not necessarily the only possible solution of the IP. Other metrics with a different monopole and other higher multipoles could in fact produce the same $\delta(z)$. An additional degeneracy can come from inverting the metric using observations along the light cone.

The metric we obtain with our inversion method can be considered an effective one which does solve the inversion problem, but it is not the only possible solution, and becomes unique only in the linear limit when the effects of different multipoles can be decoupled or in the nonlinear regime under the a-priori assumption of spherical symmetry, i.e. in absence of any higher multipole. 
This degeneracy is related to the back-reaction effect due to the non commutativity of spatial averaging  with the non linear differential operators present in  the Einstein equations \cite{Buchert:1999er}. In the case of the Friedman equations back-reaction terms arise from spatial averaging, while in our case, when performing angular averages, the Einstein tensor of the monopole of the metric and the monopole of the Einstein tensor can differ by some analogous back-reaction term. The metric obtained by inversion from the monopole of the energy momentum tensor is an effective metric which includes some of these back-reactions terms, which are related to the angular average of higher multipoles. In fact the same degeneracy happens for the FLRW metric used as an effective description of the Universe on large scales, which can correspond to several different inhomogeneous metrics which all give the same effective FLRW metric after spatial averaging.
In our case the effective metric has spherical symmetry,  several anisotropic metrics could produce the same monopole of the  Einstein tensor, corresponding to the same the same $\delta(z)$, and  solve the IP. 

The result of the inversion  should thus be considered the monopole of the effective metric corresponding to assuming isotropy when angular averaging is performed on sufficiently large scales. In our case  the homogeneity of the effective FLRW metric is replaced by the spherical symmetry of the LTB metric, which is supposed to be a well defined effective metric on sufficiently large angular scales, but on smaller scale this effective description may not be accurate.
It should be noted in fact that if local structure were highly anisotropic this notion of effective metric may not be well defined, in the same way the FLRW effective description would not be very accurate  if the Universe were highly inhomogeneous on all scales.

\subsection{Testing the accuracy of the inversion method}

\begin{figure}[ht]
 \centering
    \begin{subfigure}{0.48\textwidth}
       \includegraphics[width=\textwidth]{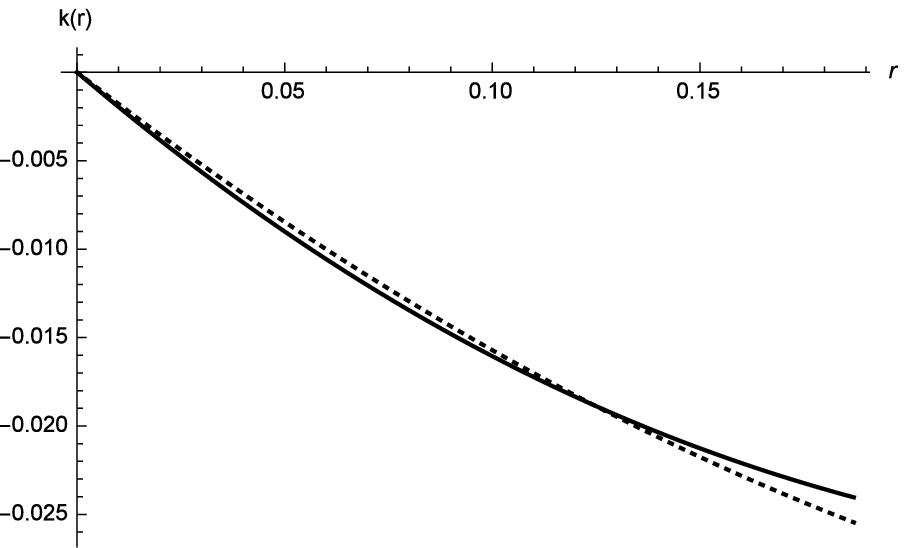}
    \end{subfigure}
    \quad
	\begin{subfigure}{0.48\textwidth}
       \includegraphics[width=\textwidth]{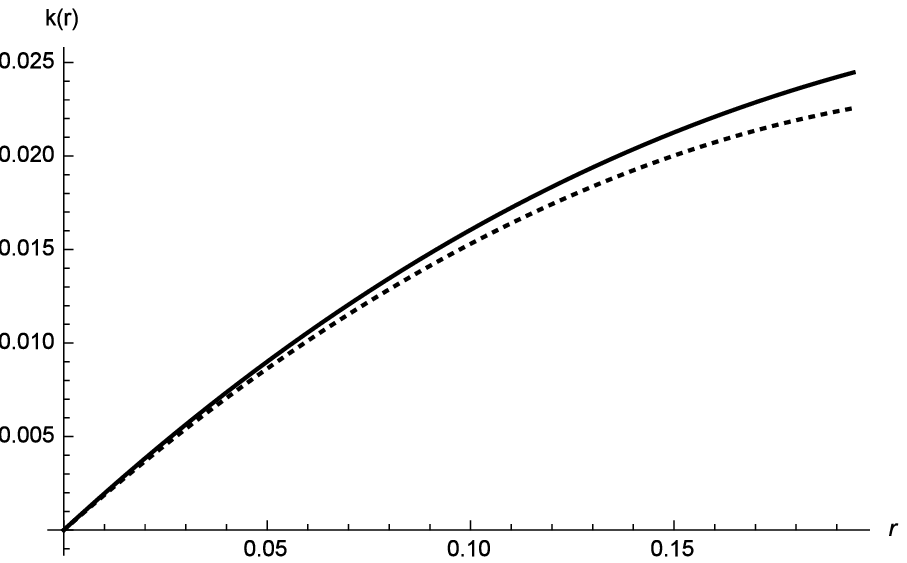}
    \end{subfigure}
 \caption{The reconstructed metric function $k(r)$  is plotted in units of  $H_0^{2}$ as a function of the radial coordinate in units of $H_0^{-1}$ for the inhomogeneities corresponding to Fig 1. The black solid line corresponds to the original $k(r)$ function 
and the black dotted line to the reconstructed one. } 
\label{IPKnp}
\end{figure}

In order to test the inversion method we compute numerically the density using the models defined in eq.(\ref{eq:57}) and then calculate the corresponding  low red-shift expansion of $\delta(z)$.  The coefficients of the reconstructed $k(r)$ are then obtained from  eq.(\ref{eq:54}).
The result of the inversion is then compared to the original $k(r)$ defined in eq.(\ref{eq:57}).
As shown in fig.(\ref{IPKnp}) at low red-shift the reconstructed  $k(r)$ is in good agreement with the numerical results. 

As shown in the previous section the perturbative calculation for $\delta(z)$ is less accurate than the analytical formula we computed in eq.(\ref{eq:51}-\ref{eq:52}). We can infer that also the pertubative solution of the inversion problem, which would consist in solving the perturbed Einstein equations to get the metric from the density contrast, will be less accurate than the analytical method we have developed. 
% With these interpolated functions we were able to obtain the coefficients $\delta_0$, $\delta_1$ and $\delta_2$ corresponding to the two numerical solutions we considered. 
% We then plotted the analytical solution for $k(r)$ and the functions defined in eq.(\ref{eq:57}) up to a small radius of $r_l=r(z_l)=r(0.1)$ as can be seen in Fig. 5. 

%If we want the inhomogeneous LTB solution to match the density profile of a homogeneous FLRW Universe then $\rho(z)=\rho_{H}(z)$ must hold, we shall call this the matching condition. This condition implies that $\rho_Z=\rho_{Z_{H}}$, then from Eqs. (44) and (49) it follows that $H_{0}^2 \Omega _{M}=H_{0H}^2 \Omega _{M_{H}}$. We then substitute in Eqs. (50) and (51), and making use of the fact that the matching condition also implies that $\rho_F=\rho_{F_{H}}$ and $\rho_S=\rho_{S_{H}}$ we can solve for $K_1$ and $K_2$ from Eqs. (45-46) and (50-51). The only possible solution is found to be $K_1=K_2=0$, which corresponds to the case where there is no inhomogeneity, thus there is no inhomogenous solution able to mimic the matter density profile in redshift space of an homogeneous FLRW Universe according to our results.  

%\section{Testing the accuracy of the formulae}

%In order to test the accuracy of the formulae we compare to the numerical solution of an inhomogeneity modeled by 

%\bea
%k(r)= 
%\eea
\section{Conclusions}

% Using an exact solution of the Einstein field equation for an observer located at the center of a spherically symmetric matter distribution we have studied the effects of the local structure due to its monopole component. 
We have derived the low-redshift expansion for the monopole of the density profile and the density contrast. At low red-shift the formulae  are in good agreement with numerical solutions and are more accurate than the linear perturbation theory approximation.  
% While the  expansion for $r(z)$ depends on the choice of the radial coordinate, the formulae for $\rho(z)$ and $\delta(z)$ do not depend on it because $\rho$ and $\delta$ are observer independent quantities which. 
% Nonetheless there is still some coordinate dependency left in the way we parameterize the inhomogeneity, this is because in the derivation of the formulae the expansion coefficients of the function $k(r)$ are used to determine the inhomogeneity profile. However our gauge choice is quite natural because in the FLRW gauge when the function $k(r)$ is a constant the radial coordinate reduces to the comoving coordinate of an FLRW solution with its curvature given by the constant value of $k$. For this reason the asymptotic behavior of the functions defined in eq.(\ref{eq:57}) to solve numerically eq.(\ref{eq:2}), (\ref{eq:4}) and (\ref{eq:5})  allowed us to defined the background quantities following the relations given in eq.(\ref{eq:59}). \\
Using these formulae we have then developed a new analytical inversion method to reconstruct the metric from the monopole  of the density contrast.
% To do this we considered the low red-shift formulae for the density constrast we derived and solved algebraically for the coefficients $K_1$ and $K_2$ in terms of the density contrast.
% We found from this solution that if the density contrast corresponds to the density contrast of a homogeneous Universe, i.e. $\delta=0$ then the only possible solution is the homogenous one. 
The inversion method could be applied to low red-shift observational data to determine the metric with a level of precision higher than the one achievable  using perturbation theory. 

In the future the formulae we obtained for the metric could  be used in the expansion of other cosmological observables to get coordinate independent formulas for these quantities in terms of the density contrast, without the need to expand the metric. It will also be interesting to develop a numerical inversion method able to reconstruct the metric beyond the regime of validity of the low red-shift expansion or to adopt other more accurate expansion techniques such as the Pad\'e approximation. It will also be interesting to compare the results of the inversion method with perturbation theory methods in the Newton gauge used in modern galaxy analysis such as those given in \cite{Bonvin:2011}. This can be achieved by re-writing the LTB metric in the Newton gauge \cite{VanAcoleyen:2008}, and it would be important to check if the perturbative treatment including the effect of peculiar velocity and light propagation is in good agreement with our results or if other  nonlinear effects can be important.

For a full reconstruction of the metric beyond the monopole contribution  other solutions of the Einstein equations could be used for the analytical approach, in order to accommodate more complex  geometries. For a general numerical inversion able to reconstruct any type of metric more sophisticated  methods in numerical relativity will be required.\\

\section*{Acknowledgments}
We thank the anonymous Referee for the suggestions to improve the manuscript and to follow up this paper with future projects about the comparison with other more accurate  perturbative approaches.

%\begin{figure}[h]
%\centering
%\includegraphics[width=0.9\textwidth]{DL_z.eps}  
%\caption{The luminosity distance DL is plotted as a function of the redshift z. The black solid line corresponds to the numerical result, the dashed line corresponds to the analytical formula in terms of $K_1$ and $K_2$, the dotdashed line is for the analytical formula in terms of $\delta_1$ and $\delta_2$, and the blue solid line corresponds to the flat FLRW model in presence of a cosmological constant and to perturbation theory result around this model, i.e., the contribution from perturbation theory is very small.}
%\label{dlz}
%\end{figure}

\appendix
\section{ Derivation of the analytical formula}
In order to obtain the formulae for the red-shift expansion of $\rho(z)$ and $\delta(z)$ we have applied several manipulations and substitutions. 
The method is based on re-expressing everything in terms of physical quantities, starting from the definitions of $a_0$ and $H_0$, which are related to $\wp$ and $\wp '$ by the equations
\begin{align}
a_0 & \equiv  (\eta_0,0) = \frac{\rho _0}{k_0 + 3 \wp _0} \,, \\
H_0 & \equiv  H(\eta_0,0) = - \frac{3 \wp' _0}{2 \rho _0} \,,
\end{align}
where 
\begin{align}
\wp _0 &= \wp (\eta _0; g_2(0),g_3(0)) \,, \\
\wp' _0 &= \frac{\partial \wp (\eta; g_2(0),g_3(0)) }{\partial \eta} |_{\eta=\eta _0} \,.
\end{align}
By inverting the previous equations we obtain the following relations
\begin{align}
\wp _0 &= \wp (\eta _0; g_2(0),g_3(0)) = \frac{\rho _0 - a_0 k_0}{3 a _0} \,,\\
\wp' _0 &= \frac{\partial \wp (\eta; g_2(0),g_3(0)) }{\partial \eta} |_{\eta=\eta _0} = - \frac{2 H_0 \rho _0}{3} \,.
\end{align}
We can then substitute the above expressions everywhere $\wp$ and $\wp '$ appear, making  the final formula only depending on physical quantities such as $H_0$. 

In order to simplify the results we have also used the Einstein equation for the LTB metric at the center $(\eta _0 ,0)$ 

\begin{equation}
1=-K_0+\Omega _M + \Omega _{\Lambda} \, ,
\end{equation}

and assumed a flat $\Lambda CDM$ for the background

\begin{equation}
1=\Omega^b _M + \Omega _{\Lambda}^b \, .
\end{equation}

\section{General formulae}

Here we give the  low red-shift formulae for the density and for  the solution to the inversion problem for the general case in which $k_0$ is different from zero. All the formulae are found using the computer algebra system provided by the Wolfram Mathematica software. We also test the accuracy of the general formulae against numerical calculations and linear perturbation theory for the density contrast. In order to do this comparisons we consider the type I$^{-}$ inhomogeneity studied in \cite{Romano:2010nc}

\begin{figure}[ht]
     \includegraphics[width=0.85\textwidth]{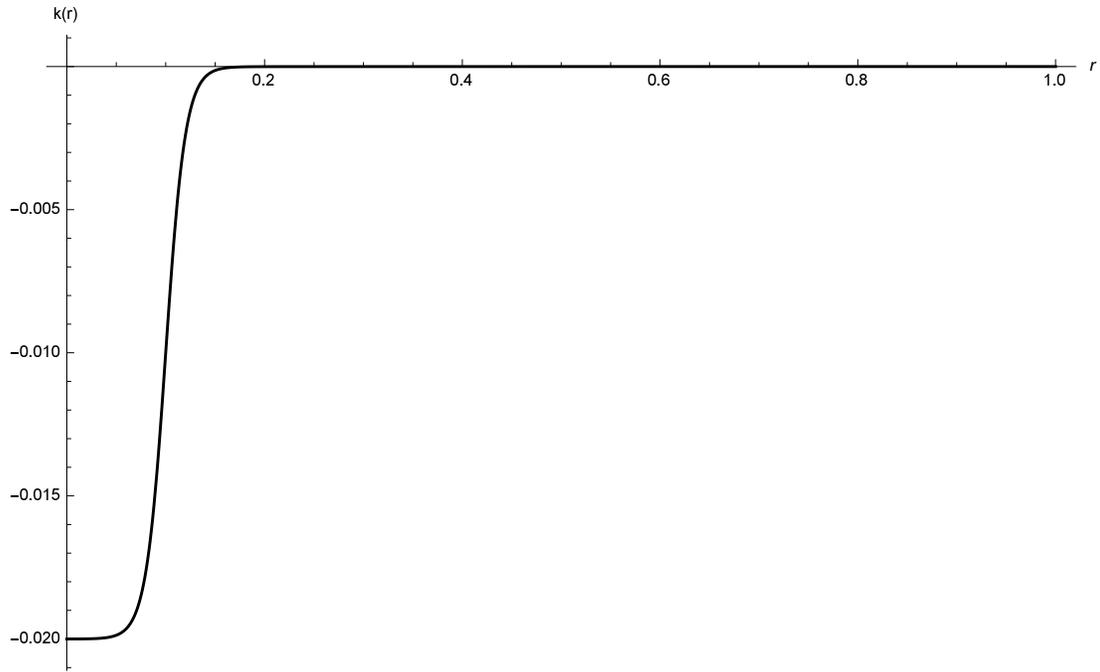}
     \caption{The function $k(r)$ corresponding to the type I$^{-}$ inhomogeneity studied in \cite{Romano:2010nc} is plotted in units of  $H_0^{2}$  as a function of the radial coordinate in units of $H_0^{-1}$.}   
\label{kng}
\end{figure}

For the density profile we have
\begin{align}
\rho(z) &= 3 H_0^2 \Omega_M + \rho_1 z + \rho_2 z^2 \, , \\  
\rho_1&= \frac{3 H_0^2 \Omega _M}{27 \Omega _{\Lambda } \Omega _M^2-4 K_0^3} \left[-4 K_0^3 \left(4 \alpha  K_1+3\right)+9 \Omega _{\Lambda } \Omega _M \left(4 K_1 \left(3 \alpha  \Omega _M+1\right)+9 \Omega _M\right)+ \right. \nonumber \\ & \quad {} \left. +24 K_1 K_0 \left(\zeta _0-\Omega
   _M\right)+4 K_1 K_0^2 \left(T_0+2\right)\right] \, ,\\ 
\rho_2 &=  \frac{3 H_0^2 \Omega _M }{4 \left(4 K_0^3-27 \Omega _M^2 \Omega _{\Lambda }\right){}^2}   \bigg[-16  \left(K_1 \left(T_0+4\right)-12\right) K_0^6+ 2 \left(T_0 \left(3 T_0+16\right) K_1^2+ \right. \nonumber \\ & \quad {} \left.   +12 \left(\left(T_0+12\right) \Omega _M-4 \left(2 T_0+\zeta _0+4\right)\right)
   K_1-40 K_2 \left(T_0+2\right)\right) K_0^5+\left(\big(8 \left(24 \zeta _0+35\right)+ \right.  \nonumber \\ & \quad {}  \left. \left.   +T_0  \left(8 \left(5 T_0+9 \zeta _0+25\right)-9 \left(T_0+16\right) \Omega _M\right)\right) K_1^2+144
   \big(\Omega _M \left(\zeta _0-2 \Omega _M-2 \Omega _{\Lambda }+9\right)+ \right. \nonumber \\ & \quad {} \left. -8 \zeta _0\big) K_1+480 K_2 \left(\Omega _M-\zeta _0\right)\right) K_0^4+12 \left(\left(18 \zeta _0^2+\left(40
   \left(T_0+3\right)-9 \left(T_0+8\right) \Omega _M\right) \zeta _0+ \right. \right. \nonumber \\ & \quad {}  \left. \left.  +\Omega _M \left(T_0 \left(12 \Omega _M+12 \Omega _{\Lambda }-43\right)-126\right)\right) K_1^2+9 \Omega _M
   \left(\left(T_0+8\right) \Omega _M-16\right) \Omega _{\Lambda } K_1+ \right. \nonumber \\ & \quad {}    -12 \Omega _M \left(5 K_2+18 \Omega _M\right) \Omega _{\Lambda }\Big) K_0^3-18 \left(2 \left(-40 \zeta _0^2+6 \Omega _M^2
   \left(-4 \zeta _0+T_0 \Omega _{\Lambda }-6\right)+ \right. \right. \nonumber \\ & \quad {}  \left. \left. +\Omega _M \left(\zeta _0 \left(9 \zeta _0+86\right)-\left(20 T_0+24 \zeta _0+75\right) \Omega _{\Lambda }+5\right)\right) K_1^2+9 \Omega _M^2
   \left(\left(T_0+12\right) \Omega _M+ \right. \right. \nonumber \\ & \quad {}   \left.  -4 \left(2 T_0+\zeta _0+4\right)\right) \Omega _{\Lambda } K_1-30 K_2 \left(T_0+2\right) \Omega _M^2 \Omega _{\Lambda }\Big) K_0^2+ \nonumber \\ & \quad {} +\left(\left(\left(6 K_0-9 \Omega _M+50\right) \alpha ^2+10 \beta \right) K_1^2+20 \alpha  K_2\right) \left(4
   K_0^3-27 \Omega _M^2 \Omega _{\Lambda }\right){}^2+  \nonumber \\ & \quad {}  +108 \Omega _M \Omega
   _{\Lambda } \left(\left(40 \zeta _0+\left(5 T_0-12 \zeta _0-28\right) \Omega _M\right) K_1^2+30 K_2 \left(\zeta _0 -\Omega _M\right)\Omega _M +\right. \nonumber \\ & \quad {} \left. +9 \Omega _M \left(8 \zeta _0+\Omega _M \left(-\zeta _0+2 \Omega _M+2 \Omega _{\Lambda
   }-9\right)\right) K_1   \right) K_0 +324 \Omega _M^2 \Omega _{\Lambda } \big(9 \left(4-\Omega _M\right) \Omega _M
   \Omega _{\Lambda } K_1 +  \nonumber \\ & \quad {} \left. +5 \left(\zeta _0-\Omega _M+2 \Omega _{\Lambda }\right) K_1^2+ 3 \Omega _M \left(5 K_2 +9 \Omega _M\right) \Omega _{\Lambda }\right)+2 \alpha  K_1 \left(4 K_0^3-27 \Omega _M^2 \Omega _{\Lambda }\right) \left(8 K_0^4+ \right. \nonumber \\ & \quad {} \left.  -2 \left(K_1 \left(3
   T_0+8\right)+6 \left(\Omega _M-8\right)\right) K_0^3+ K_1 \left(-50 \left(T_0+2\right)-36 \zeta _0+9 \left(T_0+8\right) \Omega _M\right) K_0^2+ \right. \nonumber \\ & \quad {} \left.  +6 K_1 \zeta _0 \left(9 \Omega _M-50\right) K_0 -6
   \Omega _M \left(9 \Omega _M \Omega _{\Lambda }+K_1 \left(12 \Omega _M+12 \Omega _{\Lambda }-53\right)\right) K_0+ \right. \nonumber \\ & \quad {} \left. +9 \Omega _M \left(9 \left(\Omega _M-8\right) \Omega _M   +2 K_1 \left(6 \Omega
   _M-25\right)\right) \Omega _{\Lambda }\right)\bigg] \, .
\end{align}

\begin{figure}[ht]     \includegraphics[width=0.85\textwidth]{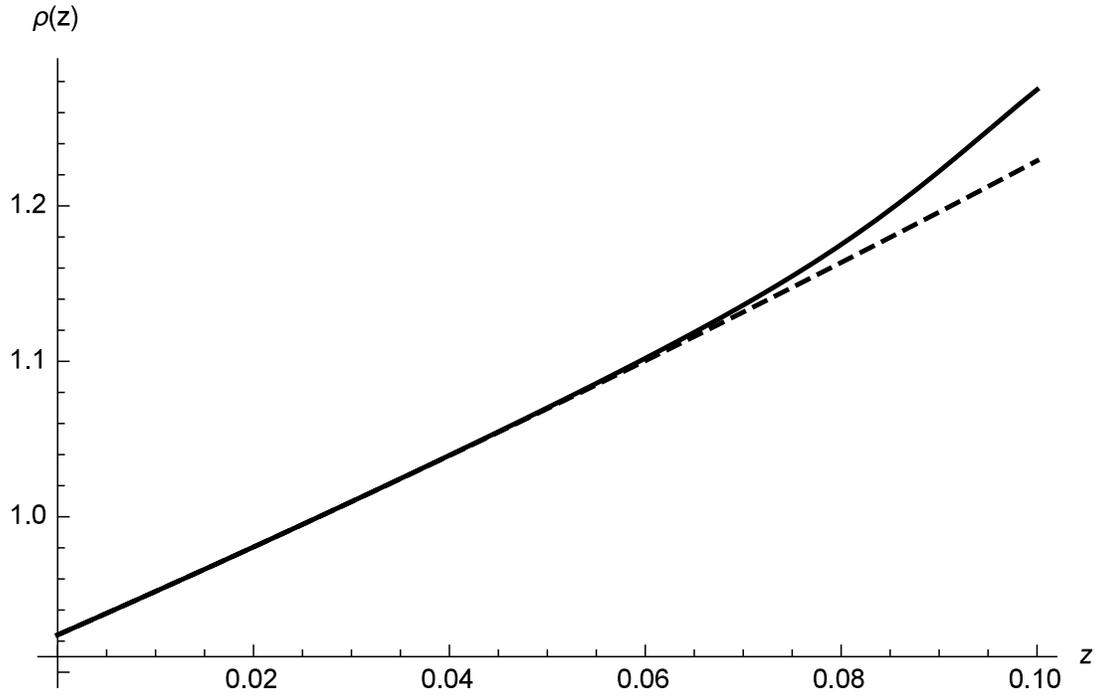}
     \caption{The density profile in units of $H_0^{2}$ is plotted as a function of redshift for the type I$^{-}$ inhomogeneity. The solid line is for the numerical calculation and the dashed line for the analytical approximation.}   
\label{rhong}
\end{figure}

%\pagebreak

The low red-shift expansion for the density contrast is
\begin{align}
\delta(z)&=\delta_0 + \delta_1 z + \delta_2 z^2 \, , \\
\delta_0&= \left(\frac{H_0}{H_0^b}\right)^2\frac{\Omega _M}{ \Omega _{M}^b}-1 \, , \\ 
\delta_1&= -\left(\frac{H_0}{H_0^b}\right)^2\frac{4 K_1 \Omega _M \left(-4 \alpha  K_0^3-6 K_0 \left(\Omega _M-\zeta _0\right)+K_0^2 \left(T_0+2\right)+9 \Omega _{\Lambda } \Omega
   _M \left(3 \alpha  \Omega _M+1\right)\right)}{ \Omega _{M}^b \left(4 K_0^3-27 \Omega _{\Lambda } \Omega _M^2\right)} \, , \\
\delta_2&= \left(\frac{H_0}{H_0^b}\right)^2 \frac{\Omega _M}{4 \Omega _{M}^b \left(4 K_0^3-27 \Omega _{\Lambda } \Omega _M^2\right){}^2} \bigg[\left(4 K_0^3-27 \Omega _{\Lambda } \Omega _M^2\right){}^2 \left(20 \alpha  K_2+K_1^2 \big(10 \beta +  \right. \nonumber \\ & \quad {} \left. \left. \left. +\alpha ^2
   \left(6 K_0-9 \Omega _M+50\right)\right)\right)+324 \Omega _{\Lambda } \Omega _M^2 \left(5 K_1^2 \left(\zeta _0+2 \Omega _{\Lambda }-\Omega
   _M\right)-9 K_1 \Omega _{\Lambda } \Omega _M^2+  \right. \right. \nonumber \\ & \quad {}  \left.  +15 K_2 \Omega _{\Lambda } \Omega _M\big)+2 \alpha  K_1 \left(4 K_0^3-27 \Omega _{\Lambda }
   \Omega _M^2\right) \left(-6 K_0 \left(K_1 \big(50 \zeta _0 + \right. \right. \right. \nonumber \\ & \quad {} \left. \left. \left. \left. +\Omega _M \left(-9 \zeta _0+12 \Omega _{\Lambda }-53\right)+12 \Omega
   _M^2\right)+9 \Omega _{\Lambda } \Omega _M^2\right)+9 \Omega _{\Lambda } \Omega _M \left(2 K_1 \left(6 \Omega _M-25\right)+9 \Omega
   _M^2\right) +  \right. \right. \nonumber \\ & \quad {} \left. +K_1 K_0^2 \left(9 \left(T_0+8\right) \Omega _M-2 \left(18 \zeta _0+25 T_0+50\right)\right)-2 K_0^3 \left(K_1 \left(3
   T_0+8\right)+6 \Omega _M\right)+8 K_0^4\bigg) + \right. \nonumber \\ & \quad {} \left. +K_0^4 \left(-144 K_1 \Omega _M \left(-\zeta _0+2 \Omega _{\Lambda }+2 \Omega _M-1\right)+480
   K_2 \left(\Omega _M-\zeta _0\right)+K_1^2 \big(8 \left(24 \zeta _0+35\right)  + \right. \right. \nonumber \\ & \quad {} \left. \left. \left. -8 T_0 \left(-9 \zeta _0+18 \Omega _M-25\right)+T_0^2 \left(40-9
   \Omega _M\right)\right)\right)+12 K_0^3 \Big(-60 K_2 \Omega _{\Lambda } \Omega _M  +  \right. \nonumber \\ & \quad {} \left. \left. +K_1^2 \left(\Omega _M \left(T_0 \left(-9 \zeta _0+12
   \Omega _{\Lambda }-43\right)-18 \left(4 \zeta _0+7\right)\right)+12 T_0 \Omega _M^2+2 \zeta _0 \left(9 \zeta _0+20 T_0+60\right)\right) +  \right. \right. \nonumber \\ & \quad {} \left. +9 K_1
   \left(T_0+8\right) \Omega _{\Lambda } \Omega _M^2\Big)-18 K_0^2 \left(2 K_1^2 \Big(-40 \zeta _0^2+6 \Omega _M^2 \left(-4 \zeta _0+T_0
   \Omega _{\Lambda }-6\right)+  \right. \right. \nonumber \\ & \quad {} \left. \left. \left. +\Omega _M \left(9 \zeta _0^2+86 \zeta _0-\Omega _{\Lambda } \left(24 \zeta _0+20 T_0+75\right)+5\right)\right)+9
   K_1 \Omega _{\Lambda } \Omega _M^2 \left(\left(T_0+12\right) \Omega _M-4 \zeta _0\right)+  \right. \right. \nonumber \\ & \quad {} \left. -30 K_2 \left(T_0+2\right) \Omega _{\Lambda } \Omega
   _M^2\bigg)+108 K_0 \Omega _{\Lambda } \Omega _M \left(9 K_1 \Omega _M^2 \left(-\zeta _0+2 \Omega _{\Lambda }+2 \Omega _M-1\right)+  \right. \right. \nonumber \\ & \quad {} \left. \left. +30 K_2
   \Omega _M \left(\zeta _0-\Omega _M\right)+K_1^2 \left(40 \zeta _0+\Omega _M \left(-12 \zeta _0+5 T_0-28\right)\right)\right)+  \right. \nonumber \\ & \quad {} \left. +K_0^5 \left(2
   K_1 \left(-48 \zeta _0+K_1 T_0 \left(3 T_0+16\right)+12 \left(T_0+12\right) \Omega _M\right)-80 K_2 \left(T_0+2\right)\right)+  \right. \nonumber \\ & \quad {} -16 K_1 K_0^6
   \left(T_0+4\right)\bigg] \, . 
\end{align}

\begin{figure}[ht] 
\includegraphics[width=0.7\textwidth]{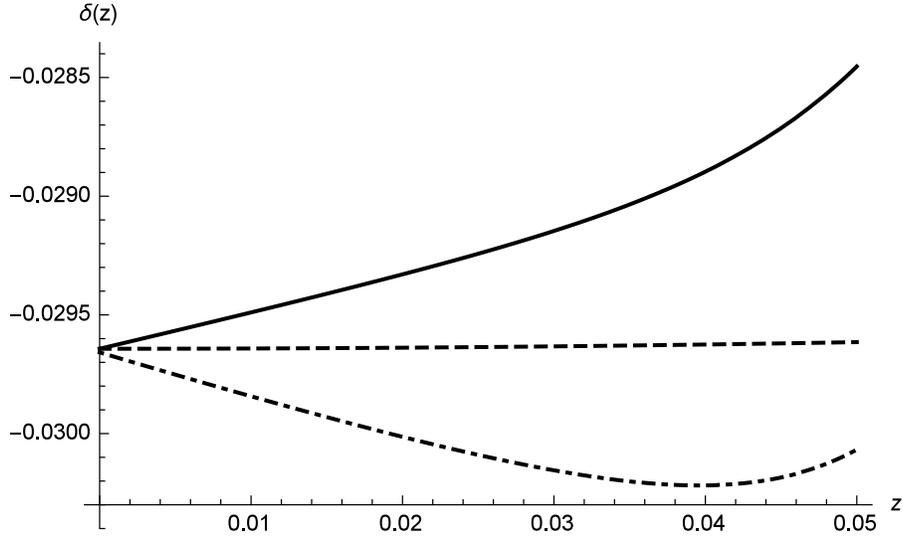}
     \caption{The density contrast is plotted as a function of redshift for the type I$^{-}$ inhomogeneity. The solid line corresponds to the numerical solution, the dashed line to the analytical formula we derived and the dot-dashed line to the perturbation theory result.}   
\label{deltarhong}
\end{figure}

\begin{figure}[ht] 
\includegraphics[width=0.65\textwidth]{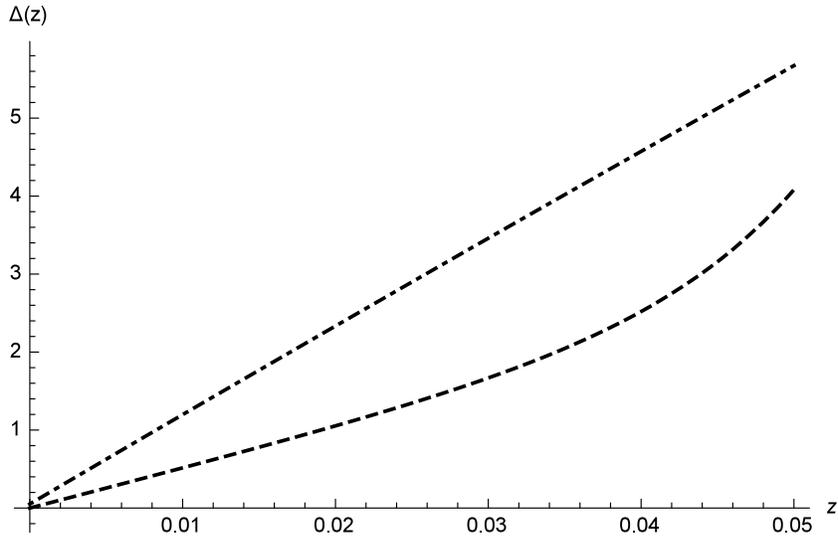}
     \caption{The relative percentual difference between different analytical approximations and the numerical calculation $\Delta(z)=100 \left(\frac{\delta^A(z)}{\delta^N(z)}-1\right)$  is plotted as a function of redshift for the type I$^{-}$ inhomogeneity. The dashed line is for the analytical formula and the dot-dashed line for the perturbation theory approximation.}   
\label{Ddeltarhong}
\end{figure}

The solution of the inversion problem is 
%\allowdisplaybreaks
\begin{align}
k(r)&= (a_0 H_0)^2 K_0 + (a_0 H_0)^3 K_1 r + (a_0 H_0)^4 K_2 r^2 \, , \\ 
K_1&= \left(\frac{H_{0}^b}{H_{0}}\right)^2 \frac{\delta _1 \Omega _{M}^b \left(27 \Omega _{\Lambda } \Omega _M^2-4 K_0^3\right)}{4 \Omega _M\left(-4 \alpha 
   K_0^3-6 K_0 \left(\Omega _M-\zeta _0\right)+K_0^2 \left(T_0+2\right)+9 \Omega _{\Lambda } \Omega _M \left(3 \alpha  \Omega
   _M+1\right)\right)} \, , \\
K_2&=  \left(\frac{H_{0}^b}{H_{0}}\right)^4 \frac{ \Omega _{M}^b }{320 \Omega _M^2 } \mathcal{A}  \Bigg[8 \Omega _M \left(4 \alpha 
   K_0^3-\left(T_0+2\right) K_0^2+6 \left(\Omega _M-\zeta _0\right) K_0+ \right. \nonumber \\ & \quad {} \left. -9 \Omega _M \left(3 \alpha  \Omega _M+1\right) \Omega _{\Lambda }\big)
   \Big(8 \alpha  \delta _1 K_0^4-2 \left(16 \alpha  \delta _2+\delta _1 \left(T_0+6 \alpha  \Omega _M+4\right)\right) K_0^3+ \right. \nonumber \\ & \quad {} \left. \left. +\left(4 \left(2
   \left(T_0+2\right) \delta _2-3 \delta _1 \zeta _0\right)+3 \left(T_0+12\right) \delta _1 \Omega _M\right) K_0^2-6 \left(3 \delta _1 \left(3
   \alpha  \Omega _{\Lambda }+2\right) \Omega _M^2+ \right. \right. \right. \nonumber \\ & \quad {} \left. \left. \left. +\left(8 \delta _2+\delta _1 \left(-3 \zeta _0+6 \Omega _{\Lambda }-3\right)\right) \Omega
   _M-8 \delta _2 \zeta _0\right) K_0+9 \Omega _M \left(9 \alpha  \delta _1 \Omega _M^2+ \right. \right. \right. \nonumber \\ & \quad {} \left. \left. +6 \left(\delta _1+4 \alpha  \delta _2\right) \Omega _M+8
   \delta _2\big) \Omega _{\Lambda }\right) \left(\frac{H_{0}}{H_{0}^b}\right)^2+ \delta _1^2 \Omega _{M}^b  \Bigg(96 \alpha ^2 K_0^7-16 \left(\left(9
   \Omega _M-50\right) \alpha ^2+ \right. \right. \nonumber \\ & \quad {} \left. \left. +\left(3 T_0+8\right) \alpha -10 \beta \big) K_0^6+\left(6 T_0^2+8 \left(\alpha  \left(9 \Omega
   _M-50\right)+4\right) T_0+ \right. \right. \right. \nonumber \\ & \quad {} \left. \left. +32 \alpha  \left(-9 \zeta _0+18 \Omega _M-25\right)\big) K_0^5+\Big(\left(40-9 \Omega _M\right) T_0^2-8 \left(-9
   \zeta _0+18 \Omega _M-25\right) T_0+ \right. \right. \nonumber \\ & \quad {} \left. \left. \left. +8 \left(-18 \alpha  \left(9 \alpha  \Omega _{\Lambda }+4\right) \Omega _M^2+6 \alpha  \left(53-12 \Omega
   _{\Lambda }\right) \Omega _M+6 \zeta _0 \left(\alpha  \left(9 \Omega _M-50\right)+4\right)+35\right)\right) K_0^4+ \right. \right. \nonumber \\ & \quad {} \left. \left. +12 \left(162 \alpha ^2
   \Omega _{\Lambda } \Omega _M^3+3 \left(T_0 \left(9 \alpha  \Omega _{\Lambda }+4\right)-12 \left(25 \alpha ^2-4 \alpha +5 \beta \right) \Omega
   _{\Lambda }\right) \Omega _M^2+ \right. \right. \right. \nonumber \\ & \quad {} \left. \left. +\left(T_0 \left(-9 \zeta _0+12 \Omega _{\Lambda }-43\right)-6 \left(12 \zeta _0+50 \alpha  \Omega _{\Lambda
   }+21\right)\right) \Omega _M+2 \zeta _0 \left(20 T_0+9 \zeta _0+60\right)\bigg) K_0^3+ \right. \right. \nonumber \\ & \quad {} \left. \left. -18 \Big(27 \alpha  \left(T_0+8\right) \Omega
   _{\Lambda } \Omega _M^3-6 \left(8 \zeta _0+\left(25 T_0 \alpha +18 \zeta _0 \alpha +50 \alpha -2 T_0\right) \Omega _{\Lambda }+12\right)
   \Omega _M^2+ \right. \right. \nonumber \\ & \quad {} \left. \left. \left.  +2 \left(9 \zeta _0^2+86 \zeta _0-\left(20 T_0+24 \zeta _0+75\right) \Omega _{\Lambda }+5\right) \Omega _M-80 \zeta _0^2\right)
   K_0^2+ \right. \right. \nonumber \\ & \quad {} \left. \left. +54 \Omega _M \Omega _{\Lambda } \left(9 \alpha  \left(9 \alpha  \Omega _{\Lambda }+8\right) \Omega _M^3+6 \alpha  \left(-9 \zeta _0+12
   \Omega _{\Lambda }-53\right) \Omega _M^2+ \right. \right. \right. \nonumber \\ & \quad {} \left. \left. +2 \left(5 T_0+150 \alpha  \zeta _0-12 \zeta _0-28\right) \Omega _M+80 \zeta _0\big) K_0-81 \Omega
   _M^2 \Omega _{\Lambda } \bigg(20 \left(\Omega _M-\zeta _0\right)+ \right. \right. \nonumber \\ & \quad {} \left. \left. +\left(81 \alpha ^2 \Omega _M^3-18 \left(25 \alpha ^2-4 \alpha +5 \beta
   \right) \Omega _M^2-300 \alpha  \Omega _M-40\right) \Omega _{\Lambda }\right)\right)\Bigg] \, ,
\end{align}

where 

\begin{equation}
\mathcal{A}= \frac{\left(4 K_0^3-27 \Omega _M^2 \Omega _{\Lambda }\right)}{ \left(-4 \alpha 
   K_0^3+\left(T_0+2\right) K_0^2-6 \left(\Omega _M-\zeta _0\right) K_0+9 \Omega _M \left(3 \alpha  \Omega _M+1\right) \Omega _{\Lambda
   }\right){}^3}
\end{equation}

\begin{figure}[ht] 
\includegraphics[width=0.85\textwidth]{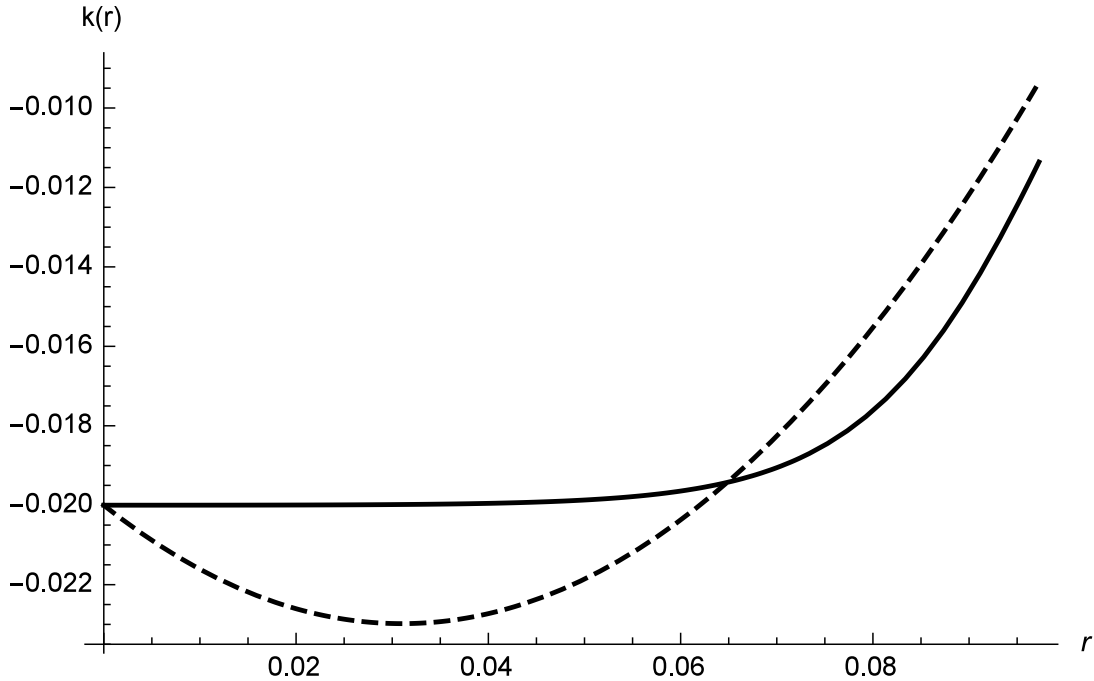}
     \caption{The reconstructed metric function $k(r)$  is plotted in units of  $H_0^{2}$ as a function of the radial coordinate in units of $H_0^{-1}$ for the type I$^{-}$ inhomogeneity. The black solid line corresponds to the original $k(r)$ function and the black dotted line to the reconstructed one.}   
\label{IPKg}
\end{figure}

\acknowledgments

\bibliographystyle{h-physrev4}
\bibliography{mybib}
\end{document}